\documentclass[journal]{jpp}

       \newcommand{\Cc}{ {\mathcal{C}} }

       \newcommand{\Hc}{ {\mathcal{H}} }
       
       \newcommand{\Jc}{ {\mathcal{J}} }

       \newcommand{\Mc}{ {\mathcal{M}} }
       
       \newcommand{\Oc}{ {\mathcal{O}} }

       \newcommand{\Uc}{ {\mathcal{U}} }

       \newcommand{\bfa}{{\mathbf a}}
       \newcommand{\bfb}{{\mathbf b}}

       \newcommand{\bfu}{{\boldsymbol u}}

       \newcommand{\bfx}{{\boldsymbol x}}

       \newcommand{\bfA}{{\boldsymbol A}}
       \newcommand{\bfB}{{\boldsymbol B}}

       \newcommand{\bfE}{{\boldsymbol E}}

       \newcommand{\bfJ}{{\boldsymbol J}}

       \newcommand{\bfM}{{\boldsymbol M}}

       \newcommand{\bfV}{{\boldsymbol V}}

       \newcommand{\bspi}{{\boldsymbol \pi}}
  \newcommand{\nn}{\nonumber}

  \newcommand{\nb}{\nabla}
  \newcommand{\hz}{\hat{z}}
  
  \newcommand{\nbvp}{\nabla_{v_\perp}}
  
  \newcommand{\bsv}{\boldsymbol{v}}
  \newcommand{\bxi}{\boldsymbol{\xi}}
  \newcommand{\bbfM}{\bar{\mathbf{M}}}
  \newcommand{\nbv}{\nabla_{\boldsymbol{v}}}
  \newcommand{\tP}{\textbf{\textsf{\textit{P}}}}
  \newcommand{\tQ}{\textbf{\textsf{\textit{Q}}}}
  \newcommand{\tA}{\textbf{\textsf{\textit{A}}}}
  \newcommand{\tB}{\textbf{\textsf{\textit{B}}}}
  \newcommand{\tI}{\textbf{\textsf{\textit{I}}}}

\usepackage{graphicx}
\usepackage{epstopdf, epsfig}
\usepackage[colorlinks]{hyperref}
\usepackage{amsfonts}
\usepackage{amssymb}
\usepackage[T1]{fontenc}
\usepackage{mathtools}
\usepackage{amsfonts}
\usepackage{dcolumn}
\usepackage{bm}
\usepackage[export]{adjustbox}
\usepackage{wrapfig}
\usepackage{lipsum}
\allowdisplaybreaks
\usepackage[titletoc,toc,title]{appendix}

\shorttitle{Hamiltonian kinetic-Hall MHD with fluid-kinetic ions}
\shortauthor{D. A. Kaltsas, G. N. Throumoulopoulos, and P. J. Morrison}

\title{Hamiltonian kinetic-Hall magnetohydrodynamics with fluid and kinetic ions in the current and pressure coupling schemes}

\author{D. A. Kaltsas\aff{1}
  \corresp{\email{d.kaltsas@uoi.gr}},
  G. N. Throumoulopoulos \aff{1}
 \and P. J. Morrison \aff{2}}

\affiliation{\aff{1}Department of Physics, University of Ioannina, Ioannina, Greece, GR 451 10.
\aff{2}Department of Physics and Institute for Fusion Studies,University of Texas, Austin, Texas 78712, USA}
\date{2021}
\begin{document}
\maketitle

\begin{abstract}
We present two generalized hybrid kinetic-Hall magnetohydrodynamics (MHD) models describing the interaction of a two-fluid bulk plasma, which consists of thermal ions and electrons, with energetic, suprathermal ion populations described by Vlasov dynamics. The dynamics of the thermal components are governed by standard fluid equations in the Hall MHD limit with the electron momentum equation providing an Ohm's law with Hall and electron pressure terms involving a gyrotropic electron pressure tensor. The coupling of the bulk, low-energy plasma with the energetic particle dynamics is accomplished through the current density (current coupling scheme; CCS) and the ion pressure tensor appearing in the momentum equation (pressure coupling scheme; PCS) in the first and the second model, respectively. The CCS is a generalization of two well-known models, because in the limit of vanishing energetic and thermal ion densities we recover the standard Hall MHD and the hybrid kinetic-ions/fluid-electron model, respectively. This provides us with the capability to study in a continuous manner the global impact of the energetic particles in a regime extending from vanishing to dominant energetic particle densities. The noncanonical Hamiltonian structures of the CCS and PCS,  which can be exploited  to study equilibrium and stability properties through the energy-Casimir variational principle, are identified. As a first application here, we derive a generalized Hall MHD Grad--Shafranov--Bernoulli system for translationally symmetric equilibria with anisotropic electron pressure and kinetic effects owing to the presence of energetic particles using the PCS. 
\end{abstract}

%
%
%
%
%

\section{Introduction}
\label{sec_I}
The presence of energetic, suprathermal particle populations, which can modify the global plasma dynamics owing to their high kinetic energies while having small density compared with that of thermal particles, is a common feature in astrophysical and fusion plasmas. For example, such energetic particles are part of the galactic cosmic rays \citep{Amato2018} originating outside the solar system and penetrate the magnetospheric plasma. Additionally, charged particles are energized and accelerated during solar flares and coronal mass ejections, owing to magnetic reconnection and shock formation \citep{Klein2017}. Furthermore, it is known that in the magnetospheric ring-current plasma, energetic particles coexist with particles of significantly lower energy that constitute the plasma bulk (e.g. see \cite{Daglis1999}). Although, the density of the thermal particles is dominant, the high-energy content of the fast particles renders even small populations capable of significantly affecting the plasma dynamics. Analogous situations, where suprathermal particles are produced and interact with a thermal plasma bulk occur in fusion experiments by external plasma heating mechanisms, such as neutral beam injection (NBI) and  ion cyclotron resonance heating (ICRH), which accelerate hydrogen isotopes and $^3He$ to energies of order 1 MeV \citep{Start1999}. Also, in a burning plasma the deuterium–tritium  (D–T) fusion reactions produce  particularly energetic alpha particles (3.5 MeV) that are expected to heat the plasma, and consequently their impact on the dynamics of burning plasmas cannot be neglected. Energetic particles in fusion experiments are responsible for the destabilization of the Alfv\'en eigenmodes (AEs) (e.g. see \cite{Wesson2011,Chen2007,Todo2019}) owing to resonant wave--particle interactions occuring 
when there exists a particle population with velocities near the phase velocity of the Alfv\'en wave. These interactions lead eventually to particle losses, which prevent the energetic particles from transferring their energy to the thermal plasma and thus deteriorate the heating efficiency.

The coexistence of a thermal or cold bulk and suprathermal populations has motivated the development of several hybrid multi-scale plasma models using fluid equations to describe the bulk plasma and Vlasov, or reduced kinetic equations (drift kinetic and gyrokinetic) to describe the energetic particle dynamics. Of course, a fully kinetic description using, for example, the Maxwel--Vlasov system, contains all the micro- and macro-physics involved in such systems. However, the hybrid fluid--kinetic description significantly reduces the computational cost because it is not required to simulate the dynamics of all particle species using kinetic equations. This makes the hybrid models important tools for performing numerical simulations and studying nonlinear dynamics because resolving all the kinetic scales of large systems with complex geometries is an extremely demanding task in terms of computational resources.

The hybrid models require a set of fluid and kinetic equations that should self-consistently describe the interaction of the plasma bulk with the energetic particles and the electromagnetic fields. For the fluid equations, the most popular choice is to consider an MHD description while for the kinetic component, one may use the Vlasov or reduced kinetic theories if the magnetic field is strong and the particle magnetic moment is an adiabatic invariant.  To couple the dynamics of the plasma components there are two main strategies: one relies on coupling through the current density, called the current coupling scheme (CCS) while the other is a pressure coupling scheme (PCS), where the kinetic effects are introduced through the pressure tensors appearing in the fluid momentum equation. The kinetic-MHD model in the PCS was introduced by \cite{Cheng1991} and the first CCS kinetic-MHD, which considered gyrokinetic particles was introduced by \cite{Park1992} to study the nonlinear behaviour of energetic particle effects.  Since then, several hybrid drift kinetic and gyrokinetic MHD models have been employed to simulate plasma dynamics containing Alfv\'en eigenmodes, using either the CCS or the PCS, for example \cite{Briguglio1995,Todo1998,Park1999} and \cite{Zhu2016}. 

The CCS and PCS variants of the drift kinetic and gyrokinetic MHD models were formulated using Hamiltonian variational principles and Euler--Poincar\'e reduction \citep{Burby2017,Close2018}, and before that, a Hamiltonian approach to the hybrid description of plasmas combining the noncanonical Poisson bracket \citep{Morrison1980,Morrison2009} of ordinary MHD with the particle bracket was developed by \cite{Tronci2010} (see also \cite{pjmTT14}). This approach resulted in Vlasov kinetic-MHD models in the CCS and the PCS.  The Hamiltonian construction of these models opened up the possibility to employ the energy-Casimir Hamiltonian variational principle \citep{Morrison1998} to derive equilibrium and stability conditions for planar kinetic-MHD using CCS and PCS in the studies by \cite{pjmTT14} and \cite{Tronci2015}. \cite{Tronci2010} also provides the derivation of a noncanonical Poisson bracket that correctly describes the dynamics of a standard hybrid model that treats the electrons as a fluid with zero inertia while retaining a Vlasov description for the ions (see e.g. \cite{Winske2003}). This model resolves the ion kinetic scales but not those of the electrons, thus saving computational resources while reproducing the structural details of the reconnection region, which may consist of thin current sheets with thickness of the order of the ion inertial length. These characteristics, render the model a popular choice for studying magnetic reconnection \citep{Hesse1994,Le2016,Cerri2017}.

Recently, a generalized, quasineutral hybrid model which contained an additional  fluid ion component while considering an arbitrary number of kinetic species, was presented by \cite{Amano2018}. This model makes no assumptions regarding the electron mass and employs the CCS for coupling the fluid and kinetic components. The generalized model of \cite{Amano2018} incorporates the quasineutral two-fluid (QNTF) description and the common hybrid approach of purely kinetic ions and fluid electrons with finite or vanishing electron inertia. This unifying framework enables the study of kinetic effects in a continuous manner, starting from the fluid description and proceeding to situations with several kinetic and thermal species.

In this paper we follow the approach of \cite{Amano2018} to derive the corresponding model in the CCS for inertia-less electrons accompanied with its Hamiltonian formulation. Starting from Hamiltonian theories, such as the Hall MHD \citep{Holm1987,Lingam2015} and the Vlasov equation, which has no collision operator,  one expects to find a Hamiltonian structure for the resulting hybrid model. The identification of this structure might be important for the construction of structure-preserving Hamiltonian algorithms that improve the stability and the fidelity of plasma simulations \citep{Morrison2017,Kraus2017}. Such a structure preserving code has been recently developed for the simulation of MHD waves that interact with energetic particles in the framework of hybrid kinetic-MHD \citep{Holderied2021}. Moreover, we obtain a second model upon considering the method of Hamiltonian construction of \cite{Tronci2010} for the PCS. This new model has a set of Casimir invariants, i.e., global constants of motion whose gradients are elements of the Poisson kernel \citep{Morrison1998}, which provide  an interesting coupling between the fluid and the kinetic components, as was the case for the corresponding Hamiltonian kinetic-MHD model studied by \cite{Tronci2015}. This coupling leads to novel equilibrium conditions upon employing the energy-Casimir variational principle \citep{Morrison1998}. An additional feature of our study is that we consider an anisotropic, gyrotropic electron pressure tensor. Electron pressure anisotropy is a rather ubiquitous feature in guide-field magnetic reconnection, caused by electron trapping in the parallel electric fields \citep{Egedal2008,Le2009,Egedal2013} and can contribute to reconnection rates even for vanishing non-diagonal elements \citep{Cassak2015}. Also, we should note that for negligible electron inertia, a gyrotropic pressure tensor can be consistently considered, as was done by \cite{Ito2007}. An analogous treatment for Amano's QNTF model would require the introduction of finite-Larmor-radius and gyroviscous effects, which is not trivial in the Hamiltonian framework. The inclusion of electron inertial effects, along with non-diagonal electron pressure elements, seems to be important in order to reproduce fully kinetic results regarding magnetic reconnection and relevant processes within the electron diffusion region \citep{Finelli2021,Munoz2018}.  The use of a Hamiltonian approach to hybrid models with finite electron inertia would open up the possibility to construct structure-preserving algorithms for simulating a large variety of phenomena where the electron inertial effects play an important role. Although this is left for future research, we note that approaches used in the past to bridge the Hall MHD and the extended MHD Hamiltonian structures could possibly be exploited here, for example, see the works of \cite{Lingam2015} and \cite{Avignon2016}. In terms of pressure non-gyrotropy, we note that the early approach in the context of reduced fluid modelling by \cite{pjmHH87} which was significantly generalized by \cite{Lingam2020}, shows that working with the Hamiltonian structure can enable  the introduction of such effects.

The rest of the paper is organized as follows: in \S 2, we present the parent equations, the method for constructing the CCS and the associated Hamiltonian structure. In \S 3, we derive the novel hybrid model in the PCS using Tronci's method and we compare it to the model that would have been obtained using a standard (non-Hamiltonian) approach. In \S 4, we derive the translationally symmetric counterpart of the PCS Hamiltonian structure and  the associated Casimir invariants. We employ also the energy-Casimir variational principle that leads to equilibrium equations and \S 5 summarizes our results.

\section{Current coupling scheme}
\label{sec_II}
\subsection{Model equations}
The starting point for deriving the model equations in the CCS is the same as that used by \cite{Tronci2010} and \cite{Amano2018}, i.e. a set of multifluid equations governing the dynamics of the thermal components accompanied by the Vlasov equation for the energetic particles, which provide self-consistent closure to Maxwell's equations:
\begin{eqnarray}
&&\partial_t n_s+\nb\cdot(n_s\bfV_s)=0\,,\label{mf_contin}\\
&&m_sn_s\left(\partial_t\bfV_s+\bfV_s\cdot\nb\bfV_s\right)=e_sn_s\left(\bfE+\bfV_s\times\bfB\right)-\nb\cdot\tP_s\,,\label{mf_momentum}\\
&&\partial_t f_p=-\bsv\cdot\nb f_p-\frac{e_p}{m_p}\left(\bfE+\bsv\times\bfB\right)\cdot\nbv f_p\,, \label{Vlasov}\\
&&\partial_t\bfB=-\nb\times\bfE\,, \label{Faraday}\\
&&\partial_t\bfE=\epsilon_0^{-1}\mu_0^{-1}\nb\times\bfB-\epsilon_0^{-1}\bfJ\,,\label{Ampere}\\
&&\nb\cdot\bfE=\sigma/\epsilon_0\,,\quad \nb\cdot\bfB=0\,, \label{Gauss}\\
&&\sigma= \sigma_f+\sigma_k\,,\quad \sigma_f=\sum_s e_s n_s\,, \quad \sigma_k=\sum_p e_p \int d^3v f_p\,,\\ \label{sigma_def}
&&\bfJ=\bfJ_f+\bfJ_k\,,\quad \bfJ_f=\sum_s e_s n_s \bfV_s \,, \quad
\bfJ_k=\sum_p e_p \int d^3v f_p \bsv \,, \label{J_def}
\end{eqnarray}
where $\tP_s$ is the pressure tensor of the thermal species $s$,  $f_p=f_p(\bfx,\bsv,t)$ is the Vlasov distribution function, i.e., the particle density in phase space \((\bfx,\bsv)\) for the particle species \(p\). In this paper we assume that the kinetic species consist of energetic ions, e.g., populations of alpha particles or resonant ions. Note that the system \eqref{mf_contin}-\eqref{J_def} is not  fully kinetic because it has been assumed that the moment hierarchy that yields the fluid equations \eqref{mf_contin} and \eqref{mf_momentum} has been truncated in view of some appropriate fluid closure that results in independent expressions for the pressure tensors $\tP_s$ of the fluid species. For a two-fluid plasma bulk the subscript \(s\) is \(s=i,e\). Note that \(\bsv\) is the microscopic particle velocity, and therefore
\begin{eqnarray}
\frac{\partial v_i}{\partial x_j}=0\,,\quad \forall i,j\,.
\end{eqnarray}
In the low-frequency limit, it is legitimate to impose quasineutrality in a strict manner, i.e. setting $\sigma=0$. The quasineutrality assumption renders the displacement current term negligible and the Gauss law redundant. In addition, when the electron inertial effects can be ignored but the electron pressure is comparable to the magnetic pressure, the electron momentum equation results in the following generalized Ohm's law
\begin{eqnarray}
\bfE=-\bfV_e\times \bfB-\frac{\nb\cdot\tP_e}{en_e}\,. \label{Ohm_original}
\end{eqnarray}
A rigorous derivation of \eqref{Ohm_original} requires an appropriate ordering of the various terms in the electron fluid momentum equation. We may employ an Alfv\'en normalization by introducing the dimensionless quantities below
\begin{align}
&\tilde{\nb}=\ell_0 \nb\,,\quad \tilde{t}=\frac{t}{\ell_0/V_A}\,,\quad \tilde{n}_s=\frac{n_s}{n_0}\,,\nn \\  
&\tilde{B}=\frac{B}{B_0}\,,\quad \tilde{\mathrm{V}}_s=\frac{V_s}{V_A}\,, \quad \tilde{P}_s=\frac{P_s}{B_0^2/\mu_0}\,, \nn\\
&\tilde{J}=\frac{J}{B_0/(\ell_0\mu_0)}\,, \quad 
\tilde{E}=\frac{E}{B_0V_A}\,,\label{Alfven_normalization}
\end{align}
where  
$V_A=B_0/\sqrt{\mu_0 m_in_0}$ is the Alfv\'en speed, and $B_0$, $n_0$ and $\ell_0$ are the characteristic magnetic field, number density and length, respectively. The electron momentum equation can then be written in the following non-dimensional form
\begin{eqnarray}
\bfE=-\bfV_e\times\bfB-d_i\frac{\nb\cdot\tP_e}{n_e}-\frac{d_e^2}{d_i}\frac{d\bfV_e}{dt}\,. \label{nondimensional_ohm}
\end{eqnarray}
Here, the tildes have been dropped and the parameters $d_e$ and $d_i$ are the relative electron and ion skin depths, respectively. Neglecting $\Oc(d_e^2)$ terms (note that for an electron-proton plasma $d_e/d_i\sim 0.023$), thus eliminating electron length scales, the electron inertial term is removed, while the electron pressure tensor survives.  Then restoring dimensions in \eqref{nondimensional_ohm}  yields Eq. \eqref{Ohm_original}. This result stems from the assumption that the electron pressure scales as the magnetic pressure, and hence it is legitimate for plasmas with high electron $\beta$. In an alternative ordering scheme, e.g., for very low $\beta$, the order of the electron pressure term would be lower than the order of the Hall term and it could thus be neglected.

As regards the form of $\tP_e$, because of the specific ordering described above, we may consider a gyrotropic electron pressure. To see this let us write the electron pressure equation as it emerges from the moment hierarchy of the Vlasov equation (e.g. see \cite{Hunana2019a})
\begin{align}
m_e\left[\partial_t \tP_e+\nb\cdot\left(\bfV_e \tQ_e\right)+\tP_e\cdot\nb \bfV_e+(\tP_e\cdot\nb \bfV_e)^\top\right]\nn\\
=e\left[\bfB\times \tP_e+(\bfB\times\tP_e)^\top\right]\,, \label{pressure_tensor_eq}
\end{align}
where the superscript $\top$ denotes transpose and $\tQ_e$ is the electron heat flux tensor. Performing the Alfv\'en normalization \eqref{Alfven_normalization}, the electron pressure equation becomes 
\begin{eqnarray}
\frac{d_e^2}{d_i}\left[\partial_t \tP_e+\nb\cdot\left(\bfV_e\tP_e + \tQ_e\right)+\tP_e\cdot\nb \bfV_e+(\tP_e\cdot\nb \bfV_e)^\top\right]\nn\\
=\left[\bfB\times \tP_e+(\bfB\times\tP_e)^\top\right]\,.
\end{eqnarray}
Neglecting electron length scales $(d_e\rightarrow 0)$, only the right-hand side of \eqref{pressure_tensor_eq} survives, and thus $\tP_e$ satisfies $\bfB\times \tP_e+(\bfB\times\tP_e)^\top=0$. A general solution of this equation is given by the gyrotropic pressure tensor
\begin{align}
\tP_e=\frac{P_{e\parallel}-P_{e\perp} }{B^2}\bfB\bfB+P_{e\perp}\tI\,,\label{pressure_tensor}
\end{align}
where 
\begin{align}
    P_{e\parallel}&=\tP_e\boldsymbol{:}\bfb\bfb\,,\nn\\
    P_{e\perp}&=\frac{1}{2}\tP_e\boldsymbol{:} (\tI-\bfb\bfb)\,.
\end{align}
Here, $\bfb:=\bfB/|\bfB|$ and $\tA\boldsymbol{:}\tB$ indicates double contraction between the second-order tensors $\tA$ and $\tB$, i.e., $\tA\boldsymbol{:}\tB=A_{ij}B_{ij}$. We consider this specific form of electron pressure throughout the rest of the paper, because it is not only legitimate in the small electron length-scale limit, but, as will be seen in the next section, facilitates also the identification of an appropriate  Hamiltonian structure. 

Now, using $\nb\times\bfB=\mu_0\bfJ$ and \eqref{J_def}, the electron velocity can be written as
\begin{eqnarray}
\bfV_e=\frac{n_i}{n_e}\bfV-\frac{1}{en_e}\left(\mu_0^{-1}\nb\times\bfB-\sum_p e_p\int d^3v \bsv f_p\right)\,, \label{v_e}
\end{eqnarray} 
where $\bfV\equiv \bfV_i$. Inserting Eq. \eqref{v_e} in the Ohm's law \eqref{Ohm_original} we obtain
\begin{eqnarray}
\bfE=-\frac{n_i}{n_e}\bfV\times\bfB+\frac{1}{en_e}\left(\mu_0^{-1}\nb\times \bfB-\bfJ_k\right)\times\bfB-\frac{\nb\cdot\tP_e}{en_e}\,. \label{GOL}
\end{eqnarray}
Note that owing to the energetic particle component, quasineutrality does not imply $n_i = n_e$, but $n_i=n_e-e^{-1}\sigma_k$. In view of \eqref{GOL}, the Faraday's law \eqref{Faraday} leads to the following induction equation
\begin{eqnarray}
\partial_t\bfB=\nb\times\left[\frac{n_i}{n_e}\bfV\times\bfB-\frac{1}{en_e}\left(\mu_0^{-1}\nb\times\bfB-\bfJ_k\right)\times\bfB+\frac{\nb\cdot\tP_e}{en_e}\right]\,. \label{induction}
\end{eqnarray}
Inserting the generalized Ohm's law \eqref{GOL} into the ion momentum equation \eqref{mf_momentum} and the Vlasov equation \eqref{Vlasov}, we obtain respectively 
\begin{eqnarray}
&\hspace{-1cm}
mn_i\left(\partial_t\bfV+\bfV\cdot\nb\bfV\right)=\left[\frac{n_i}{n_e}\sigma_k \bfV +\frac{n_i}{n_e} \left(\bfJ-\bfJ_k \right)\right]\times\bfB-\frac{n_i}{n_e}\nb\cdot\tP_e-\nb P_i\,,\label{momentum_CCS}\\
&\hspace{-2cm}
\partial_t f_p+\bsv\cdot\nb f_p+\frac{e_p}{m_p}\left\{\left[\bsv-\frac{n_i}{n_e}\bfV+\frac{1}{en_e}\left(\bfJ-\bfJ_k\right)\right]\times\bfB -\frac{\nb\cdot\tP_e}{en_e}\right\}\cdot\nb_{\bsv} f_p=0\,,\label{Vlasov_CCS}
\end{eqnarray}
where, from now on, $m\equiv m_i$. Here, we have assumed that the thermal ions have isotropic pressure. For a magnetized plasma, a consistent consideration of thermal pressure effects for the ions within the two-fluid framework, requires the inclusion of anisotropic (and in particular non-gyrotropic) pressure effects. This will be though the topic of future research. Presently, it suffices for our purposes to consider an isotropic pressure $P_i$ assuming low temperature ions because thermal ions that deviate from the isotropic pressure description can be incorporated in the kinetic component described by the Vlasov equation.

The continuity equation for the ions remains unchanged, while for the electron fluid, inserting \eqref{v_e} into  \eqref{mf_contin}, we find
\begin{eqnarray}
\partial_t n_e=-\nb\cdot(n_i\bfV)-\frac{\nb\cdot\bfJ_k}{e}\,, \label{dn_e/dt}
\end{eqnarray}
that is 
\begin{eqnarray}
e\partial_t(n_i-n_e)=\nb\cdot \bfJ_k\,, \label{charge_conserv}
\end{eqnarray}
which is an equation for electric charge conservation. The complete system of the dynamical equations consists of the ion continuity equation, the induction equation \eqref{induction}, the momentum equation \eqref{momentum_CCS}, the Vlasov equation \eqref{Vlasov_CCS}, and the electron continuity equation \eqref{dn_e/dt} or equivalently the charge conservation equation \eqref{charge_conserv}. Computing the first-order velocity moment of the Vlasov equation, one finds 
\begin{eqnarray}
 \partial_t \sigma_k+\nb\cdot \bfJ_k=0\,,
\end{eqnarray}
which, combined with \eqref{charge_conserv}, gives the local charge conservation. Thus, for  $e(n_{i}-n_{e})+\sigma_{k}=0$ at $t=0$, the quasineutrality constraint, which is invoked in various derivations in this paper, is preserved by the dynamics.

\subsection{Hamiltonian structure}
Let us now follow the procedure introduced by \cite{Tronci2010} to derive the Hamiltonian structure of the above model. The difference here is that we apply the procedure in a more general model, by considering a two-fluid bulk plasma and generalizing also for electron pressure anisotropy. The starting point for this derivation is the combination of the Hall MHD noncanonical bracket, derived by \cite{Holm1987}, with the particle Poisson bracket. This can be done by directly adding the two brackets, which are written, however, in terms of the canonical momentum density $\bar{\bfM}:=\rho\bfV+\frac{e}{m}\rho\bfA$, where $\rho:=m n_i$, and the canonical particle momenta $\bspi_p=m_p\bsv+e_p\bfA$, where $\bfA$ is the vector potential. 

Before we proceed to this construction, let us recapitulate here some basic notions of noncanonical Hamiltonian dynamics. The Lagrange--Euler map from the Lagrangian to the Eulerian description, renders the Poisson brackets explicitly dependent on the dynamical variables of the system, say $\bxi$, i.e., it has the form
\begin{eqnarray}
\{F,G\}=\langle F_{\xi_i}, \Jc_{ij}(\bxi) G_{\xi_j}\rangle\,,
\end{eqnarray}
where $F_\xi$ represents the functional derivative of $F$ with respect to $\xi$ and $\Jc(\bxi)$ is the so-called Poisson operator.

This noncanonical Poisson bracket still satisfies the antisymmetry condition and the Jacobi identity:
\begin{eqnarray}
&\{F,G\}=-\{G,F\}\,,\\
&\{F,\{G,H\}\}+\{H,\{F,G\}\}+\{G,\{H,F\}\}=0\,,
\end{eqnarray}
where $F,G,H$ are functionals that are defined on the functional phase space. Owing to the explicit dependence on the dynamical variables there exist non-trivial functionals $\Cc$ satisfying
\begin{eqnarray}
\{\Cc,F\}=0\,,\quad \forall F\,.
\end{eqnarray}
These functionals $\Cc$ are called Casimirs. The equations of motion result from the following Hamilton equations 
\begin{eqnarray}
\partial_t \xi_i=\{\xi_i,\Hc\}\,, \label{Hamilton}
\end{eqnarray}
where $\Hc$ is the Hamiltonian of the model. Obviously, the Hamiltonian is conserved in view of the antisymmetry of the Poisson bracket and the Casimirs are constants owing to their commutative property. Such Hamiltonian structures have been identified in the context of fluid mechanics and ordinary magnetohydrodynamics \citep{Morrison1998,Morrison1980,Morrison1982}, Hall MHD \citep{Holm1987,Lingam2015}, extended-MHD \citep{Abdelhamid2015,Lingam2015} and Vlasov--Poisson and Vlasov--Maxwell theories \citep{pjm80} (with a correction for Vlasov--Maxwell in  the papers by \cite{wemo1981,Marsden1982,Morrison1982} and a limitation to the correction pointed out by \cite{Morrison1982}, which was followed up more recently by \cite{pjm13,pjmH20} and then by  \cite{lasawe2019}
).    
For a more comprehensive presentation of the noncanonical Hamiltonian dynamics emerging in the Eulerian description of fluids and other continuum theories the reader is referred to the paper by \cite{Morrison1998}. For Vlasov models the interested reader can consult the paper by \cite{Morrison2009}. 

Regarding our model, the sum of the Hall MHD and the particle brackets expressed in terms of canonical momenta is given by
\begin{eqnarray}
\{F,G\}=&&\int d^3x\Big\{ \bbfM\cdot\left(G_{\bbfM}\cdot\nb F_{\bbfM}-F_{\bbfM}\cdot\nb G_{\bbfM}\right)\nn\\
&&+\rho\left(G_{\bbfM}\cdot\nb F_{\rho}-F_{\bbfM}\cdot\nb G_{\rho}\right)
-e^{-1}\left(G_{\bfA}\cdot\nb F_{n_e}-F_{\bfA}\cdot\nb G_{n_e}\right)\nn\\
&&-\frac{1}{en_e}(\nb\times\bfA)\cdot(F_{\bfA}\times G_{\bfA})+\sum_p \int d^3\bspi_p f_p  [\![F_{\bar{f}_p},G_{\bar{f}_p}]\!]_{\bspi_p}\Big\}\,, \label{Poisson_CCS_1}
\end{eqnarray}
where 
\begin{eqnarray}
[\![g,h]\!]_{\bspi_p}=\nb g \cdot \nb_{\bspi_p} h-\nb h \cdot \nb_{\bspi_p} g\,, 
\end{eqnarray}
and $\bar{f}_p$ is the distribution function expressed in terms of $\bspi_p$, i.e. $\bar{f}_p=\bar{f}_p(\bfx,\bspi_p,t)$.

The above Poisson bracket, along with the appropriate Hamiltonian, which is the sum of the Hall MHD and particle Hamiltonians expressed in terms of $\bbfM$, $\bspi$, describe correctly the dynamics of the system. However, the canonical momentum variables are not convenient because they mix up the velocity and the magnetic field potential. Ultimately, we would like to have a system of equations that treats velocity and magnetic field variables separately, i.e. like the system we derived by the standard approach in the previous subsection. To this end, following \cite{Tronci2010} and \cite{Marsden1982}, we may perform the following change of variables:  
\begin{eqnarray}
\bbfM\rightarrow \bfV\,,\quad \bar{f}_p(\bfx,\bspi_p,t)=\bar{f}_p(\bfx,m_p\bsv+ e_p\bfA,t)\rightarrow f_p(\bfx,\bsv,t)\,,
\end{eqnarray}
where 
\begin{align}
\bfV=\rho^{-1}\bbfM-\frac{e}{m}\bfA\,. \label{v_M_trans}
\end{align}
 For convenience, this change of variables will be performed in two stages. First, we can write the bracket in terms of $f_p$ and then we can complete the change transforming from $\bbfM$ to $\bfV$. For the first change we can directly use a result from \cite{Marsden1982}, that is, for any functional $F[\bfA,f_p]$ and functions $g(\bfx,\bsv)$, $h(\bfx,\bsv)$, we have $F[\bfA,f_p(\bfx,\bsv,t)]=\bar{F}[\bfA,\bar{f}_p(\bfx,\bspi_p,t)]$,  $g(\bfx,\bsv)=\bar{g}(\bfx,\bspi_p)$,  $h(\bfx,\bsv)=\bar{h}(\bfx,\bspi_p)$, and the following transformation rules must hold:
\begin{eqnarray}
[\![\bar{g},\bar{h}]\!]_{\bspi_p}=\frac{1}{m_p}[\![g,h]\!]+\frac{e_p}{m_p^2}(\nb\times\bfA)\cdot(\nbv g\times\nbv h)\,, \label{trans_1}\\
\bar{F}_{\bfA}=F_{\bfA}-\sum_p\frac{e_p}{m_p}\int d^3v f_p\nbv F_{f_p}\,,\label{trans_2}
\end{eqnarray}
where 
\begin{eqnarray}
[\![g,h]\!]:=\nb g \cdot \nbv h-\nb h \cdot \nbv g\,. 
\end{eqnarray}
Using \eqref{trans_1} and \eqref{trans_2}, the bracket \eqref{Poisson_CCS_1} takes the following form:
\begin{align}
\{F,G\}=&\int d^3x \bigg\{\bbfM\cdot\left(G_{\bbfM}\cdot\nb F_{\bbfM}-F_{\bbfM}\cdot\nb G_{\bbfM}\right)\nn\\
&+\rho\left(G_{\bbfM}\cdot\nb F_{\rho}-F_{\bbfM}\cdot\nb G_{\rho}\right)
-e^{-1}\left(G_{\bfA}\cdot\nb F_{n_e}-F_{\bfA}\cdot\nb G_{n_e}\right)\nn\\
&-\frac{1}{en_e}(\nb\times\bfA)\cdot(F_{\bfA}\times G_{\bfA})-e^{-1}\sum_p \frac{e_p}{m_p}f_p\left[\nb G_{n_e}\cdot\nbv F_{f_p}-\nb F_{n_e}\cdot\nbv G_{f_p} \right]\nn \\
&+\frac{1}{en_e}\sum_p\frac{e_p}{m_p}f_p (\nb\times\bfA)\cdot\left(\nbv F_{f_p}\times G_{\bfA}-\nbv G_{f_p}\times F_{\bfA}\right)\nn\\
&-\frac{1}{e n_e}(\nb\times \bfA)\cdot\int \int d^3vd^3v' \left(\sum_p \frac{e_p}{m_p}f_p\nbv F_{f_p}\right)\times\left(\sum_{p'} \frac{e_{p'}}{m_{p'}}f_{p'}\nb_{\bsv'} F_{f_{p'}}\right)\nn\\
&+\sum_p \int d^3v \frac{f_p}{m_p} \left[[\![F_{f_p},G_{f_p}]\!]+\frac{e_p}{m_p}(\nb\times \bfA)\cdot\left(\nbv F_{f_p}\times\nbv G_{f_p}\right) \right]\bigg\}\,. \label{Poisson_CCS_2}
\end{align}
Now, we can proceed by expressing this bracket in terms of $\bfV$ and $\bfB=\nb\times \bfA$ instead of $\bar{\bfM}$ and $\bfA$. Let us note first that from \eqref{v_M_trans}, an arbitrary variation of $\bfV$ can be written as
\begin{eqnarray}
\delta\bfV=\rho^{-1}\delta \bar{\bfM}-\rho^{-2}\bar{\bfM}\delta\rho-\frac{e}{m}\delta\bfA\,.
\end{eqnarray}
Considering a functional $\bar{F}$ of $\bar{\bfM}$, $\bfA$ and $\rho$, and then expressing it in terms of $\bfV$, $\bfB$ and $\rho$, the equality $\bar{F}[\bar{\bfM},\bfA,\rho]=F[\bfV,\bfB,\rho]$ must hold. Upon taking the first variation of this relation and using the chain rule for the functional derivatives, the following relations can be deduced:
\begin{align}
\bar{F}_{\bar{\bfM}}=&\rho^{-1}F_{\bfV}\,,\label{F_trans_1_1}\\
\bar{F}_\rho=&F_\rho-\rho^{-1}\left(\bfV+\frac{e}{m}\bfA\right)\cdot F_{\bfV}\,,\label{F_trans_1_2}\\
\bar{F}_{\bfA}=&\nb\times F_{\bfB}-\frac{e}{m}F_{\bfV}\,. \label{F_trans_1_3}
\end{align}
Substituting \eqref{v_M_trans} and \eqref{F_trans_1_1}-\eqref{F_trans_1_3} to \eqref{Poisson_CCS_2}, we find the final bracket, which reads\footnote{Note that for deriving the bracket \eqref{Poisson_CCS_3} we have made use of the vector calculus identity
$\nb \bfb\cdot \bfa=\bfa\times\nb\times\bfb+\bfa\cdot\nb \bfb\,.$}
\begin{align}
\{F,G\}=&\int d^3x\bigg\{\left(G_\bfV\cdot \nb F_\rho -F_\bfV\cdot\nb G_\rho\right)+\rho^{-1} (\nb\times\bfV)\cdot\left(F_\bfV\times G_\bfV\right)\nn\\
&+\frac{e}{m^2}\left(\frac{mn_e-\rho}{\rho n_e}\right)\bfB\cdot\left(F_\bfV\times G_\bfV\right)+m^{-1} \left(G_\bfV\cdot\nb F_{n_e}-F_\bfV\cdot\nb G_{n_e}\right)\nn\\
&+ \frac{1}{m n_e} \bfB \cdot  \left[ F_{\bfV}\times(\nb\times G_\bfB)-G_{\bfV}\times(\nb\times F_\bfB) \right]-\frac{1}{en_e}\bfB\cdot\left[(\nb\times F_\bfB)\times(\nb\times G_\bfB)\right]\nn\\
&+\frac{1}{en_e}\sum_p\frac{e_p}{m_p}\int d^3v\, f_p \bfB \cdot\Big[\nbv F_{f_p}\times(\nb\times G_\bfB)-\nbv G_{f_p}\times(\nb\times F_\bfB)\nn\\
&+\frac{e}{m}\left(\nbv G_{f_p}\times F_{\bfV}-\nbv F_{f_p}\times G_{\bfV}\right)\Big]\nn\\
&-\frac{1}{e}\sum_p \frac{e_p}{m_p}\int d^3v\, f_p \left(\nb G_{n_e}\cdot \nbv F_{f_p}-\nb F_{n_e}\cdot \nbv G_{f_p}\right)\nn\\ 
&-\frac{1}{en_e}\bfB \cdot\left( \sum_{p,p'}\frac{e_p}{m_p}\frac{e_{p'}}{m_{p'}}\int\int d^3v d^3v'\, f_p f_{p'} \nbv F_{f_p}\times \nbv G_{f_{p'}}\right) \nn\\
&+\sum_p \int d^3v\, \frac{f_p}{m_p}\left[[\![F_{f_p},G_{f_p}]\!]+\frac{e_p}{m_p}\bfB\cdot (\nbv F_{f_p}\times \nbv G_{f_p})\right] \bigg\}\,. \label{Poisson_CCS_3}
\end{align}
Observe, unlike the bracket of \eqref{Poisson_CCS_1}, \eqref{Poisson_CCS_3} is gauge invariant.
Now, having computed the bracket of our model, we write down the Hamiltonian functional, which is the direct sum of the fluid and particle Hamiltonians, i.e.
\begin{align}
    \Hc=\int d^3x \left[\rho \frac{|\bfV|^2}{2}+\rho U(\rho)+n_e\Uc_e+\frac{|\bfB|^2}{2\mu_0}\right]+\sum_p\int\int d^3x\,d^3v\, m_p \frac{v^2}{2} f_p\,. \label{Hamiltonian_CCS}
\end{align}
Here, $U(\rho)$ is the specific internal energy of the thermal ion fluid and $\Uc_e$ is some electron internal energy function. The dependence of this function is dictated by the nature of the electron pressure. In our case,  it was first shown by \cite{Morrison1982} (and subsequent work, e.g., \cite{pjmHM13}) that gyrotropic pressure tensors of the form \eqref{pressure_tensor} follow from an internal energy that depends explicitly on $n_e$ and $|\bfB|$.  If  an  electron internal energy has the form 
\begin{eqnarray}
\Uc_e=\Uc_e(n_e,|\bfB|)\,,
\end{eqnarray}
 then, the following equations give the pressure tensor components $P_{e\parallel}$ and $P_{e\perp}$:
\begin{align}
    \frac{\partial \Uc_e}{\partial n_e}&=\frac{P_{e\parallel}}{n_e^2}\,, \label{dUe_dne}\\
    \frac{\partial \Uc_e}{\partial  |\bfB|}&=\frac{P_{e\perp}-P_{e\parallel}}{n_e|\bfB|}\,;\label{dUe_dB}
\end{align}
and hence
\begin{align}
    \frac{\delta \Hc}{\delta n_e}=\Uc_e+\frac{P_{e\parallel}}{n_e}\,,\label{dH/dne}\\
    \frac{\delta \Hc}{\delta \bfB}=\mu_0^{-1}(1-\gamma)\bfB\,, \label{dH/d|B|}
\end{align}
where
$\gamma:= (P_{e\parallel}-P_{e\perp})/(B^2/\mu_0)$ is a function that measures the electron pressure anisotropy. Note that the Poisson structure \eqref{Poisson_CCS_3} is valid for barotropic ($\Uc_e=\Uc_e(n_e)$) and gyrotropic [\eqref{dUe_dne} and \eqref{dUe_dB}] electron pressure. The identification of an appropriate Hamiltonian structure for a generally non-gyrotropic electron pressure tensor will be pursued in a future work.

The equations of motion follow from \eqref{Hamilton} with the Poisson bracket \eqref{Poisson_CCS_3}, the Hamiltonian \eqref{Hamiltonian_CCS} and also using equations \eqref{dH/dne}, \eqref{dH/d|B|}. 
It can be readily seen that $\partial_t n_e=\{n_e,\Hc\}$ and $\partial_t\rho=\{\rho,\Hc\}$ are indeed equations \eqref{dn_e/dt} and the continuity equation for the fluid ions, respectively. The latter is 
\begin{eqnarray}
\partial_t\rho=-\nb\cdot(\rho\bfV)\,.
\end{eqnarray}
The remaining equations, i.e. the momentum, the induction and the Vlasov equation that stem from \eqref{Hamilton}, are
\begin{eqnarray}
&&\hspace{-15mm}\partial_t\bfV=\bfV\times\nb\times\bfV -\nb\left( h+\frac{\mathrm{v}^2}{2} \right)+\frac{e}{m}\left(1-\frac{n_i}{n_e} \right)\bfV\times\bfB\nn \\ &&\hspace{-10mm}+\frac{1}{mn_e}\left(\mu_0^{-1}\nb\times\bfB -\bfJ_k\right)\times\bfB-m^{-1}\nb\left( \Uc_e+\frac{P_{e\parallel}}{n_e}\right)-\frac{\nb\times(\gamma \bfB)}{\mu_0mn_e}\times\bfB \,,\label{dv/dt}\\
&&\hspace{-15mm}\partial_t \bfB=\nb\times\left[\frac{n_i}{n_e}\bfV\times\bfB-\frac{1}{en_e}\left(\mu_0^{-1}\nb\times\bfB-\bfJ_k\right)\times\bfB\right]\nn\\
&&\hspace{-8mm}+\mu_0^{-1}\nb\times\left[\frac{\gamma}{en_e}\left(\bfB\cdot\nb\bfB-\nb\bfB\cdot\bfB\right)+\frac{\nb\gamma\times\bfB}{en_e}\times\bfB\right]\,,\label{dB/dt}\\
&&\hspace{-15mm}\partial_t f_p=-\bsv\cdot\nb f_p-\frac{e_p}{m_p}\Big[\left(\bsv-\frac{n_i}{n_e}\bfV\right)\times\bfB\nn\\
&&\hspace{-10mm}+\frac{1}{en_e}\left(\mu_0^{-1}\nb\times\bfB-\bfJ_k\right)\times\bfB -\frac{1}{e}\nb\left(\Uc_e+\frac{P_{e\parallel}}{n_e}\right)-\frac{\nb\times(\gamma\bfB)}{\mu_0en_e}\times\bfB\Big]\cdot\nbv f_p\,. \label{df/dt}
\end{eqnarray}

It is a matter of vector calculus manipulations to prove that
\begin{eqnarray}
&&-\frac{1}{m}\nb\left(\Uc_e+\frac{P_{e\parallel}}{n_e}\right)-\frac{\nb\times(\gamma\bfB)}{\mu_0mn_e}\times\bfB=\nn\\
&&-\frac{1}{mn_e}\left[\nb P_{e\perp}+\mu_0^{-1}(\gamma \bfB\cdot\nb\bfB +\bfB\bfB\cdot\nb\gamma) \right]=-\frac{\nb\cdot \tP_e}{mn_e}\,.\label{div_tensor}
\end{eqnarray}
The last equality can be easily proven by taking the divergence of the gyrotropic pressure equation \eqref{pressure_tensor}. In light of \eqref{div_tensor}, the momentum and the Vlasov equations, \eqref{dv/dt} and  \eqref{df/dt}, respectively, which arise from the Hamiltonian formulation, are identical to \eqref{momentum_CCS} and \eqref{Vlasov_CCS}. In addition, performing some manipulations on the last term of \eqref{dB/dt} it can be seen that
\begin{eqnarray}
&&\mu_0^{-1}\nb\times\left[\frac{\gamma}{en_e}\left(\bfB\cdot\nb\bfB-\nb\bfB\cdot\bfB\right)+\frac{\nb\gamma\times\bfB}{en_e}\times\bfB\right]\nn\\
&&=\nb\times\left[\frac{\nb\cdot\tP_e}{en_e}-\frac{1}{e}\nb\Uc_e-\frac{1}{e}\nb\left(n_e\frac{\partial \Uc_e}{\partial n_e}\right)\right]=\nb\times\frac{\nb\cdot\tP_e}{en_e}\,.
\end{eqnarray}
Hence, \eqref{dB/dt} is indeed the induction equation \eqref{induction}.

\subsection{Casimir invariants}
A typical procedure to identify the Casimirs of a noncanonical Poisson bracket, e.g., \eqref{Poisson_CCS_3}, is to rearrange it in the following form:
\begin{eqnarray}
\{F,G\}=\int d^3x F_{u_i}\Jc_{ij}G_{u_j}\,,
\end{eqnarray}
where $\Jc$ is the Poisson operator associated with this bracket and then seek solutions to the following system of Casimir-determining equations:
\begin{eqnarray}
\Jc_{ij}\Cc_{u_j}=0\,,\quad i=1,...,5\,.
\end{eqnarray}
By this procedure, we find the following Casimir invariants:
\begin{eqnarray}
\Cc_1&=&\int  d^3x \rho\,,\quad
\Cc_2 = \int d^3x\, n_e\,,\\
\Cc_3&=&\frac{1}{2}\int d^3x\, \bfA\cdot\bfB\,,\\
\Cc_4&=&\frac{1}{2}\int d^3x\, \left(\bfV+\frac{e}{m}\bfA\right)\cdot\left(\nb\times\bfV+\frac{e}{m}\bfB\right)\,,\\
\Cc_p&=&\int\int d^3x d^3v\, \Lambda_p(f_p)\,.
\end{eqnarray}
This set of Casimirs, typical of magnetofluid models (e.g.\ \cite{pjmLM16}), has two helicity invariants, but is  augmented by an additional kinetic Casimir for each particle species $p$, which involves the corresponding distribution function, while there are no cross-fluid-kinetic Casimirs. Here, $\Lambda_p(f_p)$ are arbitrary functions of their respective distribution functions $f_p$.

\section{Pressure coupling scheme}
\label{sec_III}
\subsection{Transformation of the Hamiltonian CCS}
As it is well known, another method to couple the energetic and the fluid components is through the pressure tensor appearing in a center-of-mass momentum equation, rather than through the current density. This can be done upon taking the first-order fluid moment of the Vlasov equations, which govern the dynamics of the energetic species, and adding it to the momentum equation of the thermal ions. The first-order fluid moment of \eqref{Vlasov} yields
\begin{eqnarray}
m_p\partial_t (n_p\bfV_p)=e_pn_p\left(\bfE+\bfV_p\times\bfB\right)-\sum_p\nb\cdot \tilde{\tP}_p\,, \label{particle_fluid_moment}
\end{eqnarray}
where
\begin{eqnarray}
n_p=\int d^3v\, f_p\,,\quad \bfV_p=\frac{1}{n_p}\int d^3v\,\bsv f_p \,,\quad \tilde{\tP}_p=m_p\int d^3v\, \bsv\bsv f_p\,.
\end{eqnarray}
Adding \eqref{particle_fluid_moment} and \eqref{mf_momentum} for $s=i$ and using the Ohm's law \eqref{GOL}, we obtain the following momentum equation:
\begin{eqnarray}
\partial_t\bfM+m\nb\cdot(n_i \bfV\bfV)=\mu_0^{-1}(\nb\times\bfB)\times\bfB
-\nb P_i-\nb\cdot\tP_e-\sum_p\nb\cdot\tilde{\tP}_p\,, \label{gen_PCS_dM/dt}
\end{eqnarray}
where 
\begin{eqnarray}
\bfM:=mn_i\bfV+\sum_p m_pn_p \bfV_p\,. \label{c-o-m_mom_density}
\end{eqnarray}
If we replace \eqref{momentum_CCS} of the CCS model by  \eqref{gen_PCS_dM/dt}, we 
obtain a totally equivalent model coupling though the fluid and the particle components through the particle species pressure tensor terms in \eqref{gen_PCS_dM/dt}. It is logical then to expect that this reformulation of the model would have a Hamiltonian structure resulting from some change of dynamical variables. \cite{Tronci2010} showed that the reformulation of the Hamiltonian structure is effected by considering $\bfM$ instead of $\bfV$ as an independent dynamical variable. Performing this change of variables one can see how the functional derivatives with respect to the old and the new sets of dynamical variables should relate. Requiring
\begin{eqnarray}
\delta F[n_i,n_e,\bfV,\bfB,f_p]=\delta \tilde{F}[n_i,n_e,\bfM,\bfB,f_p]\,,
\end{eqnarray}
and employing the chain rule for functional derivatives the following relations can be deduced:
\begin{eqnarray}
 F_{n_e}=\tilde{F}_{n_e}\,, \quad F_{\bfB}=\tilde{F}_{\bfB}\,,\quad
F_{\bfV}=m n_i \tilde{F}_{\bfM}\,,\nn\\
F_{n_i}=\tilde{F}_{n_i}+m \bfV\cdot \tilde{F}_{\bfM}\,,\quad
F_{f_p}=\tilde{F}_{f_p}+m_p\bsv\cdot \tilde{F}_{\bfM}\,. \label{transf_v_M}
\end{eqnarray}
Substituting equations \eqref{transf_v_M} into the Poisson bracket \eqref{Poisson_CCS_3} and using the definition \eqref{c-o-m_mom_density} of $\bfM$, we find, after some algebra, the following bracket
\begin{eqnarray}
\hspace{-10mm}\{F,G\}&=&\int d^3x\,\Big\{\bfM\cdot\left(G_{\bfM}\cdot\nb F_{\bfM}-F_{\bfM}\cdot\nb G_{\bfM}\right) +n_i\left(G_{\bfM}\cdot\nb F_{n_i}-F_{\bfM}\cdot\nb G_{n_i}\right)\nn\\
\hspace{-25mm}&+&n_e\left(G_{\bfM}\cdot\nb F_{n_e}-F_{\bfM}\cdot\nb G_{n_e}\right)+\bfB\cdot\left[F_{\bfM}\times(\nb\times\bfB)-G_{\bfM}\times(\nb\times F_{\bfB})\right]\nn\\
\hspace{-25mm}&-&  \sum_p\frac{e_p}{em_p}\int d^3v\, f_p\big(\nb_{\bsv}F_{f_p}\cdot\nb G_{n_e}-\nb_{\bsv}G_{f_p}\cdot\nb F_{n_e}\big)\nn\\
\hspace{-25mm}&+&\frac{1}{en_e}\sum_p \frac{e_p}{m_p}\int d^3v f_p\bfB\cdot\left[\nb_{\bsv}F_{f_p}\times(\nb\times G_{\bfB})-\nb_{\bsv}G_{f_p}\times(\nb\times F_{\bfB})\right]\nn\\
\hspace{-25mm}&-& \frac{1}{en_e}\bfB\cdot \int \int d^3v\,d^3v'\, \sum_{p,p'}\frac{e_pe_{p'}}{m_pm_{p'}}f_p(\bsv)f_{p'}(\bsv')\nbv F_{f_p}\times\nb_{\bsv'}G_{f_{p'}}\nn\\
\hspace{-25mm}&+&\sum_p m_p^{-1}\int d^3v \, f_p \Big[[\![F_{f_p},G_{f_p}]\!]+\frac{e_p}{m_p}\bfB\cdot\left(\nbv F_{f_p}\times \nbv G_{f_p}\right)\nn\\
\hspace{-25mm}&+&m_p\left([\![F_{f_p},\bsv\cdot G_{\bfM}]\!]-[\![G_{f_p},\bsv\cdot F_{\bfM}]\!]\right)\Big]-\frac{1}{en_e}\bfB\cdot\left[(\nb\times F_{\bfB})\times(\nb\times G_{\bfB})\right]\Big\} \,. \label{Poisson_PCS_1}
\end{eqnarray}
We should also write the Hamiltonian in terms of $\bfM$ and then invoke Hamilton's equations to retrieve the dynamical system in this PCS formulation. The transformed Hamiltonian reads
\begin{eqnarray}
\Hc&=&\int d^3x \, \left[\frac{|\bfM-\boldsymbol{\Mc}|^2}{2m n_i}+ n_i \Uc_i(n_i)+n_e \Uc_e(n_e,|\bfB|)+\frac{|\bfB|^2}{2\mu_0} \right]\nn \\
&+& \sum_p \frac{m_p}{2}\int \int d^3x d^3v\, f_pv^2\,, \label{Hamiltonian_PCS_gen}  
\end{eqnarray}
where
\begin{eqnarray}
\boldsymbol{\Mc}:=\sum_p \int d^3v\, m_pf_p \bsv\,.
\end{eqnarray}
The functional derivative of $\Hc$ with respect to the new variable $\bfM$ is 
\begin{eqnarray}
\frac{\delta \Hc}{\delta \bfM}=\bfV\,.
\end{eqnarray}
Also, one should notice that the functional derivative with respect to $f_p$ is not the same as in the previous case of the CCS, but 
\begin{eqnarray}
\frac{\delta \Hc}{\delta f_p}=m_p\frac{v^2}{2}-m_p\bsv\cdot\bfV\,.
\end{eqnarray}
We can verify that Hamilton's equations are indeed the dynamical equations of the CCS model with the momentum equation being replaced by \eqref{gen_PCS_dM/dt}.

\subsection{Conventional construction of a PCS}
 The Hamiltonian system described in the previous subsection is equivalent to the CCS model of \S \ref{sec_II} because it is obtained from CCS by a mere change of variables. However, it is a commonality in the hybrid fluid--kinetic models to employ pressure coupling schemes which involve a simpler coupling of the bulk plasma with the energetic particles assuming that $n_p\ll n_i$ and $\bfM/\rho \sim \bfV$ in the dynamical equations. These assumptions result in a momentum equation governing the ion momentum density, instead of the total momentum density $\bfM$, and containing the divergence of a pressure tensor associated with the energetic particles (e.g. see \cite{Park1992,Fu2006,Kim2004,Takahashi2009}). A similar treatment in our case with a two-fluid, Hall MHD bulk, results in the following set of equations
 \begin{eqnarray}
 &&\partial_t n +\nb\cdot(n\bfV)=0\,,\\
&&\rho (\partial_t\bfV+\bfV\cdot\nb\bfV)=\bfJ\times\bfB -\sum_p\nb\cdot\tP_p-\nb\cdot\tP_e-\nb P_i\,,\\
 &&\partial_t \bfB=\nb\times\left[ \bfV\times \bfB -\frac{1}{en}\bfJ\times \bfB+\frac{1}{en}\nb\cdot\tP_e\right]\,,\\
 &&\partial_t f_p+\bsv\cdot\nb f_p+\frac{e_p}{m_p}\left [ (\bsv-\bfV)\times \bfB +\frac{1}{en}\bfJ\times\bfB-\frac{1}{en}\nb\cdot\tP_e\right]\cdot\nbv f_p=0\,. 
 \end{eqnarray}
 Here, $n=n_i=n_e$, which is the zeroth order quasineutrality condition for $n_p/n_i \ll 1$. 
 \subsection{Hamiltonian construction of a PCS}
 It has been stressed in several works \citep{Tronci2010,Tronci2014,Burby2017,Close2018} that such PC models, which are widely employed in the study of energetic particle effects on plasma stability, do not conserve some energy functional as all consistent ideal plasma models do (e.g. \cite{Morrison1980,Marsden1982,Lingam2015}). In addition, it has been shown that a Vlasov--MHD model in the pressure-coupling scheme  exhibits spurious instabilities attributed to the lack of energy conservation \citep{Tronci2014}. In contrast, a Hamiltonian variant of the model, derived by a procedure that ensures energy conservation \citep{Tronci2010}, does not contain these unphysical modes. Prompted by this observation we employ the method of \cite{Tronci2010} in our case so as to derive a simplified, yet Hamiltonian, PCS model. The idea is, instead of assuming $n_p\ll n_i$ to simplify the equations of motion, to replace the ion momentum density $\bfM-\boldsymbol{\Mc}$ by the total momentum density $\bfM$ in the Hamiltonian functional, implicitly assuming that $\bfM\approx mn_i\bfV$ on the Hamiltonian level. Hence, the term  $|\bfM-\boldsymbol{\Mc}|^2/(2m n_i)$ in \eqref{Hamiltonian_PCS_gen} becomes $|\bfM|^2/(2mn_i)$, which leads to
 \begin{eqnarray}
 \Hc=&&\int d^3x\, \left[\frac{|\bfM|^2}{2mn_i}+n_i\Uc_i(n_i)+n_e\Uc_{e}(n_e,|\bfB|)+\frac{|\bfB|^2}{2\mu_0}\right]\nn\\
 &&+\sum_p \frac{m_p}{2}\int \int d^3xd^3v\, f_p v^2\nn\\
 &&=\int d^3x\, \left[\frac{mn_i}{2}|\bfu|^2+n_i\Uc_i(n_i)+n_e\Uc_{e}(n_e,|\bfB|)+\frac{|\bfB|^2}{2\mu_0}\right]\nn\\
 &&+\sum_p \frac{m_p}{2}\int \int d^3xd^3v\, f_p v^2\,, \label{Hamiltonian_PCS_spec}
 \end{eqnarray}
 where $\bfu:=\bfM/(m n_i)$ is a weighted sum of the ion velocities.
 Note that for $n_p\ll n_i$, and if $\bfV_p$ are comparable to $\bfV$, then $\bfu=\bfV$ up to zeroth order in $n_p/n_i$.  It is convenient for us, in terms of comparing with the results of  the previous section and other studies, to write the Poisson bracket  \eqref{Poisson_PCS_1} in terms of the velocity variable $\bfu$. This is done in Appendix \ref{app_A}. The Casimirs of the bracket given in Appendix \ref{app_A} are
 \begin{eqnarray}
 \Cc_1 &=& \int d^3x\, n_i\,, \quad
 \Cc_2 = \int d^3x\, n_e\,,\\
 \Cc_3 &=& \frac{1}{2}\int d^3x\, \bfA\cdot \bfB\,,\\
 \Cc_4 &=& \int d^3x\,\bfu\cdot\left(\frac{1}{2}\nb\times\bfu+\frac{e}{m}\bfB\right)\nn\\
 &&\hspace{-10mm}-\sum_p\int d^3x\frac{m_p}{m n_i}\int d^3v f_p\bsv\cdot\left [\nb\times\left( \bfu-\frac{1}{2m n_i}\sum_{p'}m_{p'}\int d^3v' f_{p'}\bsv'\right)+\frac{e}{m}\bfB\right]\,,\\
 \Cc_p &=& \int d^3x d^3v\, \Lambda_p(f_p)\,.
 \end{eqnarray}
The equations of motion that arise from Hamilton's equations with the Poisson bracket \eqref{PCS_Poisson_u} (Appendix \ref{app_A}) and the Hamiltonian \eqref{Hamiltonian_PCS_spec}, are
\begin{eqnarray}
&&\partial_t n_i=-\nb\cdot(n_i\bfu)\,, \label{PCS_dni/dt} \\
&&\partial_t n_e=-\nb\cdot(n_e \bfu)-\frac{1}{e}\nb\cdot\bfJ_k\,,\label{PCS_dne/dt}\\
&&\partial_t \bfB=\nb\times\left[\bfu\times\bfB -\frac{1}{en_e}\left(\mu_0^{-1}\nb\times\bfB-\bfJ_k\right)\times\bfB+\frac{1}{en_e}\nb\cdot\tP_e\right]\,,\\
&&\partial_t \bfu=\bfu\times\nb\times\bfu-\nb( h_i+u^2/2) -\rho^{-1}\nb\cdot\tP_e\nn\\
&&\hspace{45mm}-\rho^{-1}\sum_p \nb\cdot\tilde{\tP}_p+\mu_0^{-1}\rho^{-1}(\nb\times\bfB)\times\bfB\,,\\
&&\partial_t f_p=-(\bsv+\bfu)\cdot\nb f_p-\frac{e_p}{m_p}\Big[\bsv\times\bfB+\frac{1}{en_e}(\mu_0^{-1}\nb\times\bfB-\bfJ_k)\times\bfB\nn\\
&&\hspace{50mm}-\frac{1}{en_e}\nb\cdot\tP_e\Big]\cdot\nb_{v}f_p+\nbv f_p\cdot\nb\bfu\cdot \bsv\,, \label{PCS_Vlasov}
\end{eqnarray}
where $\rho=m n_i$ and $h_i=\frac{1}{m}\frac{d}{d n_i}\left[n_i \Uc_i(n_i)\right]$ is the specific enthalpy of the thermal ion fluid. Computing the first-order velocity moment of the Vlasov equation \eqref{PCS_Vlasov}, we find 
\begin{eqnarray}
\partial_t\sigma_k=-\nb\cdot(\sigma_k\bfu)-\nb\cdot\bfJ_k\,,
\end{eqnarray}
which, combined with \eqref{PCS_dni/dt} and \eqref{PCS_dne/dt}, yields
\begin{eqnarray}
\partial_t \sigma=-\nb\cdot(\sigma\bfu)\,, \label{charge_continuity}
\end{eqnarray}
which is a continuity equation implying global but not local charge conservation. However, if the plasma is initially quasineutral, i.e., $\sigma=0$ at $t=0$, then quasineutrality is ensured by \eqref{charge_continuity} $\forall t>0$. 

\section{Translationally symmetric formulation and energy-Casimir equilibria}
\subsection{Translationally symmetric formulation}
In this section we consider simplified dynamics with all dynamical variables being invariant along a fixed straight axis in physical space. In this translationally symmetric case the magnetic field and macroscopic velocity can be expressed in terms of five scalar Clebsch potentials. Note that the distribution functions, although translationally symmetric in space, still depend on all three microscopic velocity coordinates. Translationally symmetric models facilitate computer simulations, because the dependency on one spatial coordinate is dropped and can be considered good approximations whenever a strong guiding magnetic field directed in a fixed direction is present.  Using Cartesian coordinates $(x,y,z)$ in physical space and assuming invariance along the  $z$ axis, the velocity and the magnetic field can be written as
\begin{eqnarray}
\bfu &=& u_z(x,y,t)\hz +\nb\chi(x,y,t)\times \hz+\nb\Upsilon(x,y,t)\,, \label{ts_decomp_1}\\
\bfB &=& B_z(x,y,t)\hz+\nb\psi(x,y,t)\times\hz\,.\label{ts_decomp_2}
\end{eqnarray}
To obtain a Hamiltonian formulation in this symmetric case we need to translate this field decomposition in a decomposition of the vector functional derivatives in terms of functional derivatives with respect to the scalar fields \(u_z,B_z,\chi,\psi,\Upsilon\). The process of deriving these relations has been described several times before, e.g., by \cite{Andreussi2010,Kaltsas2017,Grasso2017} and \cite{Kaltsas2018}. Following the same procedure here, we find 
\begin{eqnarray}
F_{\bfu}=F_{u_z}\hat{z}+\nb F_\Omega \times\hat{z}-\nb F_{w}\,, \label{F_v} \\
F_{\bfB}=F_{B_z}\hat{z}-\nb\left(\Delta^{-1}F_\psi\right)\times\hat{z}\,, \label{F_Bz}
\end{eqnarray}
where $\Delta^{-1}$ is the inverse Laplacian operator, $\Omega:=-\Delta \chi\hat{z}$ and  $w:=\Delta \Upsilon$. The translationally symmetric Hall MHD bracket is known from \cite{Kaltsas2017,Grasso2017}, and hence we have to compute the translationally symmetric cross-kinetic-Hall MHD terms and the term accounting for electron fluid thermodynamics. This is done in Appendix \ref{app_B} upon substituting \eqref{F_v} and \eqref{F_Bz} in the bracket \eqref{PCS_Poisson_u} of Appendix \ref{app_A}. The resulting translationally symmetric bracket is 
\begin{eqnarray}
\hspace{-10mm}\{F,G\}_{TS}&=&\int d^2x \bigg\{ F_\rho \Delta G_w-G_\rho\Delta F_w\nn\\
&&\hspace{-10mm}+\frac{n_e}{\rho}\left([F_\Omega,G_{n_e}]-[G_\Omega,F_{n_e}]+\nb F_w\cdot\nb G_{n_e}-\nb G_w\cdot \nb F_{n_e}\right)\nn\\
&&\hspace{-10mm}+\rho^{-1} \Omega \left([F_\Omega,G_\Omega]+[F_w,G_w]+\nb F_w \cdot \nb G_\Omega -\nb F_\Omega\cdot\nb G_w\right)\nn\\
&&\hspace{-10mm}+u_z\big([F_\Omega,\rho^{-1}G_{u_z}]-[G_\Omega,\rho^{-1}F_{u_z}]+\nb(\rho^{-1}G_{u_z})\cdot\nb F_w\nn\\
&&\hspace{-10mm}-\nb(\rho^{-1}F_{u_z})\cdot\nb G_w+\rho^{-1}F_\Upsilon G_{u_z}-\rho^{-1}G_\Upsilon F_{u_z}\big)\nn\\
&&\hspace{-10mm}+\psi \Big([F_\Omega,\rho^{-1}G_\psi]-[G_\Omega,\rho^{-1} F_\psi]+[F_{B_z},\rho^{-1}G_{u_z}]-[G_{B_z},\rho^{-1}F_{u_z}]\nn\\
&&\hspace{-10mm}+\nb F_w \cdot \nb (\rho^{-1} G_\psi)-\nb G_w \cdot \nb (\rho^{-1} F_\psi)+\rho^{-1}F_\Upsilon G_\psi -\rho^{-1} G_\Upsilon F_\psi \Big) \nn\\
&&\hspace{-10mm}+\rho^{-1}B_z\left([F_\Omega,G_{B_z}]-[G_\Omega,F_{B_z}]+\nb F_w\cdot \nb G_{B_z}-\nb G_w\cdot\nb F_{B_z} \right)\nn\\
&&\hspace{-10mm}+\frac{1}{e}\psi\left([G_{B_z},n_{e}^{-1}F_\psi]-[F_{B_z},n_e^{-1}G_\psi]\right)-\frac{1}{en_e}B_z[F_{B_z},G_{B_z}]\nn\\
&&\hspace{-10mm}-\frac{1}{e}\sum_p\frac{e_p}{m_p}\int d^3v \big[n_e^{-1}B_z\nbvp f_p \cdot \left(G_{f_p} \nb F_{B_z}-F_{f_p}\nb G_{B_z}\right)\nn\\
&&\hspace{-10mm}+\psi([n_e^{-1}F_{f_p}\partial_{v_z}f_p,G_{B_z}]-[n_e^{-1}G_{f_p}\partial_{v_z}f_p,F_{B_z}]\nn\\
&&\hspace{-10mm}+\nb\cdot(n_e^{-1}F_\psi G_{f_p}\nbvp f_p)-\nb\cdot(n_e^{-1}G_\psi F_{f_p}\nbvp f_p))\big]\nn\\
&&\hspace{-10mm}+\frac{1}{e}\sum_p\frac{e_p}{m_p}\int d^3v\, (\nbvp f_p)\cdot\left(F_{f_p}\nb G_{n_e}-G_{f_p}\nb F_{n_e}\right)\nn\\
&&\hspace{-10mm}-\frac{1}{en_e}\sum_{p,p'}\frac{e_p}{m_p}\frac{e_{p'}}{m_{p'}}\int \int d^3v d^3v' f_p f_{p'}\Big[B_z\langle F_{f_p},G_{f_{p'}}\rangle\nn\\
&&\hspace{-10mm}-\nb\psi\cdot\Big(\partial_{v_z}F_{f_p}\nb_{v_{\perp}'}G_{f_{p'}}-\partial_{v_z'}G_{f_p'}\nb_{v_{\perp}}F_{f_{p}}\Big)\Big]
\nn\\
&&\hspace{-10mm}+\sum_p\int d^3v \frac{f_p}{m_p}\Big[[\![F_{f_p},G_{f_p}]\!]_{\perp}+\frac{e_p}{m_p}\Big(B_z\langle F_{f_p},G_{f_p} \rangle \nn\\
&&\hspace{-10mm}+\nb\psi \cdot\left[\partial_{v_z}G_{f_p}\nbvp F_{f_p}-\partial_{v_z}F_{f_p}\nbvp G_{f_p}\right] \Big)\nn\\
&&\hspace{-10mm}+m_p\Big([\![F_{f_p},\rho^{-1}\left(v_zG_{u_z}+\hat{z}\cdot \bsv_\perp \times\nb G_{\Omega}-\bsv_\perp \cdot \nb G_w \right)]\!]_{\perp}\nn\\
&&\hspace{-10mm} -[\![G_{f_p},\rho^{-1}\left(v_zF_{u_z}+\hat{z}\cdot \bsv_\perp \times\nb F_{\Omega}-\bsv_\perp \cdot \nb F_w \right)]\!]_{\perp}\Big)\Big]\bigg\}\,. \label{Poisson_ts}
\end{eqnarray}
Note that we have introduced a new bracket notation, namely %
\begin{eqnarray}
[a,b]&:=&(\nb a\times\nb b)\cdot\hat{z}\,,\\
\langle a,b\rangle &:=&(\nbvp a\times \nbvp b)\cdot\hat{z}\,,\\
\left[\![a,b]\!\right]_{\perp}&:=&\nb_\perp a\cdot\nbvp b-\nb_\perp b\cdot \nbvp a\,.
\end{eqnarray}
The translationally symmetric Hamiltonian reads as follows
\begin{eqnarray}
\Hc &=& \int d^3x \Big[\frac{1}{2}\rho \left(u_z^2+|\nb\chi|^2+2[\Upsilon,\chi]+|\nb \Upsilon|^2\right)\nn\\
&&\hspace{10mm}+\frac{B_z^2}{2\mu_0}+\frac{|\nb\psi|^2}{2\mu_0}+\rho U_i(\rho)+n_e\Uc_e(n_e,B_z,|\nb\psi|)\Big]\nn\\
&&\hspace{10mm}+\sum_p\int d^3xd^3v\, \frac{1}{2}m_p f_p v^2\,, \label{ts_Hamiltonian}
\end{eqnarray}
and the functional derivatives of $\Hc$ with respect to the two scalars associated with the magnetic field are
\begin{eqnarray}
\frac{\delta\Hc}{\delta B_z} &=& \mu_0^{-1}(1-\gamma)B_z\,,\\
\frac{\delta \Hc}{\delta\psi} &=& -\nb\cdot\left[\mu_0^{-1}(1-\gamma)\nb\psi\right]\,.
\end{eqnarray}
Having these expressions and also computing the functional derivatives with respect to the remaining scalars, one can derive the translationally symmetric dynamical equations, and in addition, equilibrium and stability conditions. For equilibrium and stability analysis one has to identify the families of Casimir invariants $\Cc$ that span the non-trivial null space of the Poisson bracket \eqref{Poisson_ts}. From the Casimir determining equations, stemming from the requirement  $\mathfrak{C}_{\xi_i}=0$, where $\mathfrak{C}_{\xi_i}$ appear in the following rearrangement of \eqref{Poisson_ts}: 
\begin{eqnarray}
\{F,G\}=\int d^3x\, \Big\{F_\rho \mathfrak{C}_\rho+F_{n_e} \mathfrak{C}_{n_e}+F_{B_z} \mathfrak{C}_{B_z}+F_\psi \mathfrak{C}_\psi+F_\chi \mathfrak{C}_\chi+F_\Upsilon \mathfrak{C}_\Upsilon\nn\\
+\sum_p\int d^3v\,F_{f_p} \mathfrak{C}_{f_p} \Big\}
\end{eqnarray}
we were able to identify the following families of Casimirs
\begin{eqnarray}
\Cc_1 &=& \int d^3x\, n_e N(\psi) \,,\label{ts_Casimir_1}\\
\Cc_2 &=& \int d^3x\, \rho K\left(\varphi\right)\,,\label{ts_Casimir_2}\\
\Cc_3 &=& \int d^3x\, B_z F(\psi)\,,\label{ts_Casimir_3}\\
\Cc_4 &=& \int d^3x\, \Big(\Omega+\frac{e}{m}B_z-\sum_p m_p \int d^3v [\rho^{-1}f_p,\bsv\cdot\bfx]\Big)G(\varphi)\,,\label{ts_Casimir_4}\\
\Cc_{p} &=& \int \int d^3xd^3v\,\Lambda_p(f_p)\,,\label{ts_Casimir_5}
\end{eqnarray}
where $N,K,F,G$ and $\Lambda_p$ are arbitrary functions of their respective arguments and
\begin{eqnarray}
\varphi:=u_z+\frac{e}{m}\psi-\rho^{-1}\sum_pm_p\int d^3v f_p v_z\,,
\end{eqnarray}
is a generalized ion stream function, modified owing to the presence of kinetic particle species. For \(f_p\rightarrow 0\), this stream function and the entire set of Casimir invariants \eqref{ts_Casimir_1}-\eqref{ts_Casimir_5}, reduce to their translationally symmetric Hall MHD counterparts \citep{Kaltsas2017}. In this fluid limit, quasineutrality implies $n_e=n_i$, because $n_p=\int d^3v\,f_p=0$.

\subsection{Energy-Casimir equilibria}
Owing to spatial symmetry, the Casimirs \eqref{ts_Casimir_1}--\eqref{ts_Casimir_5} constitute infinite families of invariants, thereby allowing for the derivation of sufficient stability conditions by the energy-Casimir method, which has been used for hybrid kinetic-MHD models by \cite{Tronci2015} and for the extended and Hall MHD models by \cite{Kaltsas2020}. A preliminary step though is the definition of the stationary state that serves as the initial condition for the dynamics. The set of equilibrium equations are derived in the first step of the energy-Casimir method by setting the first-order variation of the extended Hamiltonian equal to zero. In the MHD and extended MHD models, this procedure resulted in a system of Grad--Shafranov--Bernoulli (GSB) equations. Here, we derive the corresponding GSB system for translationally symmetric hybrid kinetic-Hall MHD in the PCS. 
The energy-Casimir functional, or extended Hamiltonian, is given by
\begin{eqnarray}
\mathfrak{F}=\Hc-\sum_{i=1}^{4}\Cc_i-\sum_p \Cc_p\,,\label{EC_functional}
\end{eqnarray}
where $\Hc$ is given by \eqref{ts_Hamiltonian} and $\Cc$ are the Casimirs \eqref{ts_Casimir_1}-\eqref{ts_Casimir_5}. The first variation of \eqref{EC_functional} can be written as 
\begin{eqnarray}
\delta\mathfrak{F} &=& \int d^3x\,\Big(\mathfrak{R}_1\delta n_e+\mathfrak{R}_2 \delta\rho+\mathfrak{R}_3 \delta u_z+\mathfrak{R}_4 \delta \chi+\mathfrak{R}_5 \delta \Upsilon\nn\\
&&\hspace{15mm}+\mathfrak{R}_6 \delta B_z+\mathfrak{R}_7 \delta\psi +\sum_p\int d^3v\, \mathfrak{R}_p\delta f_p   \Big)\,.
\end{eqnarray}
Assuming independent, arbitrary variations of the scalar dynamical variables, the requirement $\delta\mathfrak{F}=0$ is equivalent to $\mathfrak{R}_i=0$, $i=1,...,7\,$ and $\mathfrak{R}_p=0$, $\forall p$. These equations lead to the following equilibrium conditions
\begin{eqnarray}
&&\hspace{-5mm}\Uc_e+n_{e}^{-1}P_{e_{\parallel}}-N(\psi)=0\,, \label{dne}\\
&&\hspace{-5mm}h_i(\rho)+\frac{u^2}{2} -K(\varphi)-\rho^{-1}K'(\varphi)\sum_pm_p\int d^3v\, f_p v_z\nn \\
&&\hspace{-5mm}-\rho^{-2}\left(\Omega+\frac{e}{m}B_z -\sum_p m_p \int d^3v\, [\rho^{-1}f_p,\bsv\cdot\bfx] \right)G'(\varphi) \sum_{p'}m_{p'}\int d^3v\, f_{p'}v_z\nn \label{drho}\\
&&\hspace{-5mm}-\rho^{-2}\sum_p m_p \int d^3v\, f_{p}[\bsv\cdot\bfx,G]=0\,,\\
&&\hspace{-5mm}-\nb\cdot\left(\rho \nb\Upsilon \right)+[\chi,\rho]=0\,, \label{dUpsilon}\\
&&\hspace{-5mm}-\nb\cdot(\rho \nb\chi)+[\rho,\Upsilon]+\Delta G(\varphi)=0\,,\label{dchi}\\
&&\hspace{-5mm}\rho u_z-\rho K'(\varphi)-\left(\Omega+\frac{e}{m}B_z-\sum_p m_p\int d^3v [\rho^{-1}f_p,\bsv\cdot\bfx]\right)G'(\varphi)=0\,,\label{duz}\\
&&\hspace{-5mm}\mu_0^{-1}\left(1-\gamma\right)B_z-F(\psi)-\frac{e}{m}G(\varphi)=0\, \label{dBz}\\
&&\hspace{-5mm}\mu_0^{-1}\nb\cdot\left[(1-\gamma)\nb\psi\right]+n_e N'(\psi)+B_zF'(\psi)+\frac{e}{m}\rho K'(\varphi)\nn\\
&&\hspace{-5mm}+\frac{e}{m} \left(\Omega+\frac{e}{m}B_z-\sum_p m_p\int d^3v\, [\rho^{-1}f_p,\bsv\cdot\bfx]\right)G'(\varphi)=0\,,\label{dpsi}\\
&&\hspace{-5mm}\frac{1}{2}m_pv^2-\Lambda_p'(f_p)+m_pv_z\bigg[K'(\varphi)\nn\\
&&\hspace{-5mm}+\rho^{-1}\left(\Omega+\frac{e}{m}B_z-\sum_p m_p\int d^3v\, [\rho^{-1}f_p,\bsv\cdot\bfx]\right)G'(\varphi)\bigg]\nn\\
&&\hspace{-5mm}+\rho^{-1}m_p[\bsv\cdot\bfx,G(\varphi)]=0\,.\label{dfp}
\end{eqnarray}
Equation \eqref{dne} provides a relation between $P_{e_\parallel}$ and the variables $n_e$, $\psi$ and $B_z$. Upon combining \eqref{dne} with \eqref{dUe_dB} we find an equation for $P_{e_\perp}$ also. This one reads as follows 
\begin{eqnarray}
\hspace{-10mm}P_{e_\perp}=n_e\left[|\bfB|\frac{\partial \Uc_e}{\partial |\bfB|}+N(\psi)-\Uc_(n_e,|\bfB|)\right]\,. \label{Pe_perp}
\end{eqnarray}
In terms of $B_z$ and $|\nb\psi|$, \eqref{Pe_perp} becomes
\begin{eqnarray}
\hspace{-10mm}P_{e_\perp}=n_e\left[|\bfB|^2\left(\frac{1}{B_z}\frac{\partial \Uc_e}{\partial B_Z}+\frac{1}{|\nb\psi|}\frac{\partial \Uc_e}{\partial |\nb\psi|}\right)+N(\psi)-\Uc_e(n_e,B_z,|\nb\psi|)\right]\,.
\end{eqnarray}
Therefore, to fully define an equilibrium state, we need an equation of state for the electron fluid, i.e., $\Uc_e=\Uc_e(n_e,B_z,|\nb\psi|)$. For example, such equations, which are associated with collisionless magnetic reconnection, are provided by \cite{Le2009}. 

From \eqref{dBz}, we readily obtain an equation that relates $B_z$ with $\psi$, $\varphi$ and $\gamma$,
\begin{eqnarray}
B_z=\mu_0\frac{F(\psi)+(e/m)G(\varphi)}{1-\gamma}\,. \label{B_z}
\end{eqnarray}
Equation \eqref{B_z} is in general an implicit equation for determining $B_z$ because the anisotropy function, $\gamma$, depends on $B_z$ as well, and hence  $B_z$ appears nonlinearly in \eqref{dBz}. We could possibly make $\gamma$ independent of $B_z$ under some special assumption, i.e., selecting the functional dependence of $\Uc_{e}$ on $B_z$ in a particular manner.

Now, by multiplying \eqref{dfp} with $f_p$, integrating over the velocity space and summing over the particle species, we obtain the following equation
\begin{eqnarray}
\sum_p\int d^3v \frac{m_p}{2}f_p v^2+\sum_p K'(\varphi) \int d^3v m_p f_p v_z-\sum_p\int d^3v f_p \Lambda_p'(f_p) \nn\\
+\rho^{-1}\left(\Omega +\frac{e}{m} B_z -\sum_p m_p \int d^3v  [\rho^{-1}f_p,\bsv\cdot\bfx] \right)G'(\varphi)\sum_p \int d^3v m_p f_p v_z\nn\\
+\sum_pm_p \rho^{-1} \int d^3v f_p [\bsv\cdot\bfx,G(\varphi)]=0\,. \label{dfp_inter}
\end{eqnarray}
Combining equations \eqref{drho} and \eqref{dfp_inter}, we find a Bernoulli equation of the form
\begin{eqnarray}
\rho h(\rho)=\rho K(\varphi)-\frac{1}{2}\rho u^2-\sum_p\left[\int d^3v \, f_p \Lambda_p'(f_p)-\int d^3v \frac{m_p}{2}f_p v^2\right]\,. \label{Bernoulli}
\end{eqnarray}
The two Grad--Shafranov equations, one for the thermal ions and one for the electron fluid, stem from  \eqref{duz} and \eqref{dpsi}, respectively, using  \eqref{dUpsilon}, \eqref{dchi} and \eqref{dBz}, along with the definition of $\varphi$. These GS equations are
\begin{eqnarray}
&&\hspace{-15mm}G'(\varphi)\nb\cdot\left(\frac{G'}{\rho}\nb\varphi\right)+\rho\left(\varphi-\frac{e}{m}\psi\right)-\rho K'(\varphi)-\mu_0\frac{e}{m}\frac{F(\psi)+\frac{e}{m}G(\varphi)}{1-\gamma}G'(\varphi)\nn\\
&&\hspace{25mm}+\sum_p m_p \int d^3v \left(f_p v_z+G'[\rho^{-1}f_p,\bsv\cdot\bfx]\right)=0\,, \label{ion_GS}\\
&&\hspace{-15mm}\mu_0^{-1}\nb\cdot\left[(1-\gamma)\nb\psi\right]+n_e N'(\psi)+\mu_0\frac{F(\psi)+\frac{e}{m}G(\varphi)}{1-\gamma}F'(\psi)\nn\\
&&\hspace{25mm}+\frac{e}{m}\rho\left(\varphi-\frac{e}{m}\psi+\rho^{-1}\sum_p m_p \int d^3v f_p v_z\right)=0\,. \label{electron_GS}
\end{eqnarray}
Note that in \eqref{electron_GS}, the anisotropy parameter $\gamma$ appears inside the differential operator, and hence this partial differential equation might not be always elliptic \citep{Ito2007} as is the case for an isotropic electron pressure. Finally, we need a set of equations for determining the equilibrium distribution functions $f_{p,e}$. Using  \eqref{duz},  \eqref{dfp} is simplified to
\begin{eqnarray}
\Lambda_p'(f_p)=\frac{1}{2}m_p v^2+m_p \bsv\cdot\bfu\,, \label{Lambda_prime}
\end{eqnarray}
where we have used 
\begin{eqnarray}
\bfu_\perp=\rho^{-1}\nb G\times\hz\,,
\end{eqnarray}
which can be deduced from \eqref{dchi} and from the fact that $[\bsv\cdot\bfx,G]=(\nb G\times \hz)\cdot\bsv$. For invertible functions $\Lambda_p$ we take solutions to \eqref{Lambda_prime} of the form 
\begin{eqnarray}
f_{p,e}=f_{p,e}\left(\frac{1}{2} \left(|\bsv+\bfu|^2-|\bfu|^2\right)\right)\,,  \label{dist_fun_1}
\end{eqnarray}
as was the case in the paper by \cite{pjmTT14} for the planar kinetic-MHD PCS model.  Here, the subscript $e$ denotes equilibrium distribution functions. With \eqref{Lambda_prime}, the Bernoulli equation becomes
\begin{eqnarray}
\rho h(\rho)=\rho K(\varphi)-\frac{1}{2}\rho u^2-\sum_p m_p\int d^3v f_p\bsv\cdot\bfu\,. \label{Bernoulli_2}
\end{eqnarray} 
Assuming a special functional form for $f_{p,e}$ we can, in principle, compute the velocity space integrals appearing in \eqref{ion_GS}, \eqref{electron_GS} and \eqref{Bernoulli_2} to obtain a Hall MHD GSB system, modified by the presence of energetic particles. As a final remark, note that the Hall-MHD GSB equations with anisotropic electron pressure are retrieved from \eqref{ion_GS} \eqref{electron_GS}, \eqref{Bernoulli_2}, in the limit $f_p\rightarrow 0$, as can be seen upon comparing with variants of this system obtained by \cite{Kaltsas2017,Kaltsas2018} for isotropic electron pressure.


\section{Conclusions}
We constructed two kinetic-Hall MHD models with fluid and kinetic ions and fluid electrons, neglecting electron length scales. The coupling of the kinetic and the fluid components is effected through the current density in the first case (current coupling scheme) and through the pressure tensors of the particle species in the second (pressure coupling scheme). Moreover, we consider a gyrotropic electron pressure tensor, which is legitimate in the Hall MHD limit, i.e., when neglecting electron length scales. This description, allowing for both fluid and kinetic ions, bridges ordinary Hall MHD, which can be recovered in the limit of vanishing kinetic ion population, and the most common hybrid kinetic-ion/fluid-electron model often used in hybrid simulations. The basic structure of the Hall MHD dynamics is retained while enriched by kinetic effects which are consistently described by the Vlasov equation. The Hamiltonian structures of the models are identified using a method introduced by \cite{Tronci2010} and, in addition, a translationally symmetric description of the PCS model is obtained along with the corresponding Hamiltonian structure and the associated Casimir invariants. These functionals are deployed in an energy-Casimir variational principle that leads to a generalized Grad--Shafranov--Bernoulli system of equilibrium equations. We note that alternative hybrid models with non-gyrotropic electron pressure, could have been obtained if we had considered alternative extended MHD descriptions, e.g., that presented by \cite{Tronci2013} where the electron mean flow inertia is neglected in the Lagrangian of the parent kinetic theory. The use of extended MHD models with finite electron inertia and non-gyrotropic electron pressure tensors for describing the bulk plasma, and also, further investigations regarding the construction of specific equilibria and the derivation of sufficient stability criteria, will be the subject of a future work.

\section*{Acknowledgements}
The authors would like to thank the anonymous reviewers whose comments helped to improve the manuscript. PJM warmly acknowledges the hospitality of the Numerical Plasma Physics Division of Max Planck IPP, Garching, Germany, where a portion of this research was done. 

\section*{Funding}
This work   was   carried   out   within   the   framework   of   the EUROfusion Consortium and has received funding from (a) the Euratom research and training program 2014–2018 and 2019–2020 under Grant Agreement No. 633053 and (b) the National Program for  the  Controlled  Thermonuclear  Fusion,  Hellenic  Republic.  The views and opinions expressed herein do not necessarily reflect those of the European Commission. PJM was supported by the DOE Office of Fusion Energy Sciences, under DE-FG02-04ER-54742 and a Forschungspreis from the Alexander von Humboldt Foundation. 

\section*{Declaration of interests}
The authors report no conflict of interest.
\appendix

\section{PCS Poisson bracket in terms of \(\bfu\)}
\label{app_A}
To express the bracket \eqref{Poisson_PCS_1} in terms of $\bfu=\bfM/(m n_i)$, we should express the functional derivatives with respect to the new set of dynamical variables. One can prove that only the functional derivatives with respect to $\bfM$ and $n_i$ will change according to
\begin{eqnarray}
F_\bfM=\frac{\tilde{F}_\bfu}{m n_i}\,, \quad F_{n_i}=\tilde{F}_{n_i}-n_i^{-1}\bfu\cdot \tilde{F}_\bfu\,.
\end{eqnarray}
Substituting these relations into \eqref{Poisson_PCS_1} and employing some vector identities we find the following bracket
\begin{eqnarray}
&&\hspace{-15mm}\{F,G\}=\int d^3x\,\bigg\{\rho^{-1} (\nb\times\bfu)\cdot(F_\bfu\times G_\bfu)+(G_\bfu\cdot\nb F_\rho - F_\bfu\cdot \nb G_\rho)\nn\\
&&\hspace{-5mm}+\frac{n_e}{\rho}(G_\bfu\cdot \nb F_{n_e}-F_\bfu\cdot \nb G_{n_e})+\rho^{-1} \bfB\cdot \big[F_\bfu\times(\nb\times G_\bfB)\nn\\
&&\hspace{-5mm}-G_\bfu\times(\nb\times F_\bfB) \big]-\frac{1}{en_e}\bfB\cdot\left[(\nb\times F_\bfB)\times(\nb\times G_\bfB)\right]\nn\\
&&\hspace{-5mm}- \sum_p\frac{e_p}{em_p}\int d^3v\, f_p\big(\nb_{\bsv}F_{f_p}\cdot\nb G_{n_e}-\nb_{\bsv}G_{f_p}\cdot\nb F_{n_e}\big)\nn\\
\hspace{-25mm}&+&\frac{1}{en_e}\sum_p \frac{e_p}{m_p}\int d^3v f_p\bfB\cdot\left[\nb_{\bsv}F_{f_p}\times(\nb\times G_{\bfB})-\nb_{\bsv}G_{f_p}\times(\nb\times F_{\bfB})\right]\nn\\
&&\hspace{-5mm}- \frac{1}{en_e}\bfB\cdot \int \int d^3v\,d^3v'\, \sum_{p,p'}\frac{e_pe_{p'}}{m_pm_{p'}}f_p(\bsv)f_{p'}(\bsv')\nbv F_{f_p}\times\nb_{\bsv'}G_{f_{p'}}\nn\\
&&\hspace{-5mm}+\sum_p m_p^{-1}\int d^3v \, f_p \Big[[\![F_{f_p},G_{f_p}]\!]+\frac{e_p}{m_p}\bfB\cdot\left(\nbv F_{f_p}\times \nbv G_{f_p}\right)\nn\\
&&\hspace{-5mm}+m_p\left([\![F_{f_p},\rho^{-1}\bsv\cdot G_{\bfu}]\!]-[\![G_{f_p},\rho^{-1}\bsv\cdot F_{\bfu}]\!]\right)\Big]\bigg\}\,. \label{PCS_Poisson_u}
\end{eqnarray}
Note that in the limit $f_p,n_e\rightarrow 0$, \eqref{PCS_Poisson_u} becomes the Hall MHD bracket of \cite{Lingam2015}.

\section{Translationally symmetric Poisson bracket in the PCS}
\label{app_B}
The translationally symmetric counterpart of the Hall MHD bracket has been derived in detail by \cite{Kaltsas2017}. For this reason, here we provide only some details on the hybrid and purely kinetic terms under the assumption of translational symmetry in physical space, i.e., all dynamical variables are independent of the coordinate $z$.  The generic hybrid/kinetic terms are listed below:
\begin{eqnarray}
&&\{F,G\}_{h_1}=\sum_p \frac{e_p}{em_p}\int d^3x \int d^3v f_p \left(\nbv F_{f_p}\cdot \nb G_{n_e}-\nbv G_{f_p}\cdot \nb F_{n_e}\right)\,,\label{h1}\\
&&\{F,G\}_{h_2}=\frac{1}{en_e}\sum_p \frac{e_p}{m_p}\int d^3x\int d^3v\, f_p \bfB\cdot\big[\nbv F_{f_p}\times (\nb\times G_{\bfB})\nn\\
&&\hspace{20mm}-\nbv G_{f_p}\times (\nb\times F_{\bfB}) \big]\,,\label{h2}\\
&&\{F,G\}_{h_3}=-\frac{1}{en_e}\sum_{p,p'}\frac{e_pe_{p'}}{m_pm_{p'}}\int d^3x \int \int d^3v d^3v' f_p f_{p'}\bfB\cdot(\nbv F_{f_p}\times \nb_{\bsv'}G_{f_{p'}})\,,\label{h3}\\
&&\{F,G\}_{h_4}=\sum_p m_{p}^{-1}\int d^3x \int d^3v \, f_p \big\{ [\![F_{f_p},G_{f_p}]\!]+\frac{e_p}{m_p}\bfB\cdot(\nbv F_{f_p}\times \nbv G_{f_p})\nn\\
&&\hspace{20mm}+m_p\left([\![F_{f_p},\bsv\cdot G_\bfM]\!]-[\![G_{f_p},\bsv\cdot F_{\bfM}]\!]\right)\big\}\,.\label{h4}
\end{eqnarray}
The $\{F,G\}_{h_1}$ term remains essentially the same in the translationally symmetric case, except that the gradient in the velocity space is now a gradient with respect to $\bsv_\perp$, i.e., perpendicular to the $\hat{z}$ component of the microscopic velocity, because of the dot product with the gradient in physical space which has no $z$-component owing to the translational symmetry. For \(\{F,G\}_{h_2}\), we need the curl of $F_\bfB$, which is given by
\begin{eqnarray}
\nb\times F_{\bfB}=F_\psi\hat{z}+\nb F_{B_z}\times\hat{z}\,.
\end{eqnarray}
Using simple vector algebra identities we carry out the following calculation:
\begin{eqnarray}
&&\bfB\cdot\left[\nbv F_{f_p}\times (\nb\times G_{\bfB})\right]\nn\\
&&=(B_z\hz+\nb\psi\times\hz )\cdot\left[(\partial_{v_z}F_{f_p}\hz+\nbvp F_{f_p})\times(G_\psi \hz+\nb G_{B_z}\times \hz)\right]\nn\\
&&=\partial_{v_z}F_{f_p}[G_{B_z},\psi]+G_\psi \nb\psi \cdot \nbvp F_{f_p}-B_z\nb G_{B_z}\cdot\nbvp F_{f_p}\,,
\end{eqnarray}
where $[a,b]:=(\partial_x a) (\partial_y b)-(\partial_x b)(\partial_y a)$. Substituting into \eqref{h2} and performing integrations by parts, we find
\begin{eqnarray}
\{F,G\}_{h_2}&=&-\frac{1}{e}\sum_p\frac{e_p}{m_p}\int d^3x\int d^3v\Big[\psi\Big([n_{e}^{-1}F_{f_p}\partial_{v_z}f_p,G_{B_z}]\nn\\
&-&[n_{e}^{-1}G_{f_p}\partial_{v_z}f_p,F_{B_z}]+\nb\cdot(n_{e}^{-1}F_{\psi}G_{f_p}\nbvp f_p)\nn\\
&-&\nb\cdot(n_{e}^{-1}G_{\psi}F_{f_p}\nbvp f_p)\Big)+\frac{B_z}{n_e}\nbvp f_p\left(G_{f_p}\nb F_{B_z}-F_{f_p}\nb G_{B_z}\right)\Big]\,. 
\end{eqnarray}
For the third term we need to compute the triple product $\bfB\cdot(\nbv F_{f_p}\times \nb_{\bsv'} G_{f_{p'}})$,
\begin{eqnarray}
&&\bfB\cdot(\nbv F_{f_p}\times \nb_{\bsv'} G_{f_{p'}})\nn\\
&&=B_z\langle F_{f_p},G_{f_{p'}}\rangle+\nb\psi\cdot(\partial_{v_z'}G_{f_{p'}}\nbvp F_{f_p}-\partial_{v_z}F_{f_{p}}\nb_{v_\perp'} G_{f_{p'}})\,. \label{triple_product}
\end{eqnarray}
Therefore,
\begin{eqnarray}
\{F,G\}_{h_3}&=&-\frac{1}{en_e}\sum_{p,p'}\frac{e_pe_{p'}}{m_pm_{p'}}\int \int d^3v d^3v'\, f_{p}f_{p'}\big\{ B_z\langle F_{f_p},G_{f_{p'}}\rangle\nn\\
&&+\nb\psi\cdot(\partial_{v_z'}G_{f_{p'}}\nbvp F_{f_p}-\partial_{v_z}F_{f_{p}}\nb_{v_\perp'} G_{f_{p'}})\big\}\,.
\end{eqnarray}
Finally, for the bracket \( \{ F, G,\}_{h_4}\), we use again \eqref{triple_product} in conjunction with the following relation:
\begin{eqnarray}
F_{\bfM}=\rho^{-1}(F_{u_z}\hz+\nb F_\Omega\times \hz-\nb F_w)\,,
\end{eqnarray}
to arrive at
\begin{eqnarray}
\{F,G\}_{h_4}&=&\int d^3x \sum_p \frac{1}{m_p}\int d^3v\, f_p\Big\{ [\![F_{f_p},G_{f_p}]\!]_{\perp}\nn\\
&&+\frac{e_p}{m_p}\left[\langle F_{f_p},G_{f_p}\rangle+ \nb\psi\cdot\left(\partial_{v_z}G_{f_p}\nbvp F_{f_p}-\partial_{v_z}F_{f_p}\nbvp G_{f_p}\right) \right]\nn\\
&&+m_p\big([\![F_{f_p},\rho^{-1}\left(v_z G_{u_z}+\bsv_\perp\cdot\nb G_\Omega\times\hz-\bsv_\perp\cdot\nb G_w \right)]\!]_{\perp}\nn\\
&&-[\![G_{f_p},\rho^{-1}\left(v_z F_{u_z}+\bsv_\perp\cdot\nb F_\Omega\times\hz-\bsv_\perp\cdot\nb F_w \right)]\!]_{\perp}\big)\Big\}\,,
\end{eqnarray}
with the understanding that \([\![f,g]\!]_{\perp}=\nb_\perp f\cdot\nbvp g-\nb_\perp g\cdot\nbvp f\).

Adding these sub-brackets to the bracket for the translationally symmetric Hall MHD \citep{Kaltsas2017}, we form the complete translationally symmetric hybrid bracket for the pressure coupling scheme \eqref{Poisson_ts}. 

\section*{}
\bibliographystyle{jpp}
\medskip
\bibliography{biblio.bib}

\end{document}